\documentclass[twocolumn,superscriptaddress,showpacs,amsmath,amssymb,aps,pra,floatfix]{revtex4-1}
\usepackage{graphicx}
\usepackage{bm}
\usepackage{amssymb}
\usepackage{amsmath}
\usepackage{amsthm}
\usepackage{amsfonts}
\usepackage{mathrsfs}
\usepackage{hyperref}

\begin{document}

\title{Local oscillator fluctuation opens a loophole for Eve in practical continuous-variable quantum-key-distribution systems}

\author{Xiang-Chun Ma}\affiliation{Department of Physics, National University of Defense Technology, Changsha 410073, People's Republic of China}

\author{Shi-Hai Sun}\affiliation{Department of Physics, National University of Defense Technology, Changsha 410073, People's Republic of China}

\author{Mu-Sheng Jiang}\affiliation{Department of Physics, National University of Defense Technology, Changsha 410073, People's Republic of China}

\author{Lin-Mei Liang}\email{nmliang@nudt.edu.cn}\affiliation{Department of Physics, National University of Defense Technology, Changsha 410073, People's Republic of China}\affiliation{State Key Laboratory of High Performance Computing, National University of Defense Technology, Changsha 410073, People's Republic of China}

\begin{abstract}
We consider the security of practical continuous-variable quantum key distribution implementation with the local oscillator (LO) fluctuating in time, which opens a loophole for Eve to intercept the secret key. We show that Eve can simulate this fluctuation to hide her Gaussian collective attack by reducing the intensity of the LO. Numerical simulations demonstrate that, if Bob does not monitor the LO intensity and does not scale his measurements with the instantaneous intensity values of LO, the secret key rate will be compromised severely.
\end{abstract}

\maketitle

\section{\label{sec:Intro}Introduction}
Continuous-variable quantum-key distribution (CVQKD), as an unconditionally secure communication scheme between two legitimate parties Alice and Bob, has achieved advanced improvements in theoretical analysis and experimental implementation in recent years \cite{Wee12R,Gro03N,Lev09N,Lev09L,Xua09}. Practical implementation systems, such as fiber-based Gaussian-modulated \cite{Lod07,Qi07,Fos09,Jou12} and discrete-modulated \cite{Zha09,She10} coherent-state protocol QKD systems over tens of kilometers, have been demonstrated in a few groups. The unconditional security of such systems with prepare-and-measure (PM) implementation has been confirmed by the security analysis of the equivalent entanglement-based (EB) scheme \cite{Gro05,Gar06,Nav06}.

However, the traditional security analysis of the EB scheme of CVQKD just includes the signal beam and not the local oscillator (LO), which is an auxiliary light beam used as a reference to define the phase of the signal state and is necessary for balanced homodyne detection. This will leave some security loopholes for Eve because LO is also unfortunately within Eve's manipulating domain. The necessity of monitoring LO intensity for the security proofs in discrete QKD protocols embedded in continuous variables has been discussed \cite{Has08}. Moreover, in \cite{YM11}, the excess noise caused by imperfect subtraction of balanced homodyne detector (BHD) in the presence of LO intensity fluctuations has been noted and quantified with a formulation. However, in the practical implementation of CVQKD, shot noise scaling with LO power measured before keys distribution is still assumed to keep constant if the fluctuations of LO intensity are small. And in this circumstance, pulses with large fluctuation are just discarded as shown in \cite{YM11}. Unfortunately, this will give Eve some advantages in exploiting the fluctuation of LO intensity.

In this paper, we first describe Bob's measurements under this fluctuation of LO intensity, and propose an attacking scheme exploiting this fluctuation. We consider the security of practical CVQKD implementation under this attack and calculate the secret key rate with and without Bob monitoring the LO for reverse and direct reconciliation protocol. And then, we give a qualitative analysis about the effect of this LO intensity fluctuation on the secret key rate Alice and Bob hold. We find that the fluctuation of LO could compromise the secret keys severely if Bob does not scale his measurements with the instantaneous LO intensity values. Finally, we briefly discuss the accurate monitoring of LO intensity to confirm the security of the practical implementation of CVQKD.

\section{\label{sec:Loia}Local oscillator intensity fluctuation and attack}
Generally, in practical systems of CVQKD, the local oscillator intensity is always monitored by splitting a small part with a beam splitter, and pulses with large LO intensity fluctuation are discarded too. However, even with such monitoring, we do not yet clearly understand how fluctuation, in particular small fluctuation, affects the secret key rate. To confirm that the secret key rate obtained by Alice and Bob is unconditionally secure, in what follows, we will analyze the effects of this fluctuation on the secret key rate only, and do not consider the imperfect measurement of BHD due to incomplete subtraction of it in the presence of LO intensity fluctuations, which has been discussed in \cite{YM11}.

Ideally, with a strong LO, a perfect pulsed BHD measuring a weak signal whose encodings are $X_S\in\{Q_S,P_S\}$ will output the results\cite{Ray95},
\begin{equation}\label{eq:x0}
x_\theta=k|\alpha_{\text{LO}}|(Q_{\text{in}}\cos\theta+P_{\text{in}}\sin\theta),
\end{equation}
where \textit{k} is a proportional constant of BHD, $\alpha_{\text{LO}}$ is the amplitude of LO, $\theta$ is the relative phase between the signal and LO except for the signal's initial modulation phase. So scaling with LO power or shot noise, the results can be recast as
\begin{equation}\label{eq:X0}
\hat{X}_O=\hat{X}_{\text{in}}=X_S+\hat{X}_N,
\end{equation}
with $\theta$ in Eq.~(\ref{eq:x0}) is 0 or $\pi/2$. Here the quadratures $Q$ and $P$ are defined as $\hat{X}_{\text{in}}\in\{\hat{Q}_{\text{in}},\hat{P}_{\text{in}}\}$ and $\hat{X}_N\in\{\hat{Q}_N,\hat{P}_N\}$, where $\hat{X}_N$ is the quadrature of the vacuum state.

However, in a practical system, the LO intensity fluctuates in time during key distribution. With a proportional coefficient $\eta>0$, practical LO intensity can be described as $|\alpha'_{\text{LO}}|^2=\eta|\alpha_{\text{LO}}|^2$, where $\alpha_{\text{LO}}$ is the initial amplitude of LO used by normalization and its value is calibrated before key distribution by Alice and Bob. If we do not monitor LO or quantify its fluctuation \cite{foot}, especially just let the outputs of BHD scale with the initial intensity or power of LO, the outputs then read
\begin{equation}\label{eq:X'0}
\hat{X}_O^{'}=\sqrt{\eta}\hat{X}_O.
\end{equation}
Unfortunately, this fluctuation will open a loophole for Eve, as we will see in the following sections.

\begin{figure}[h]
 \scalebox{1}{\includegraphics[width=\columnwidth]{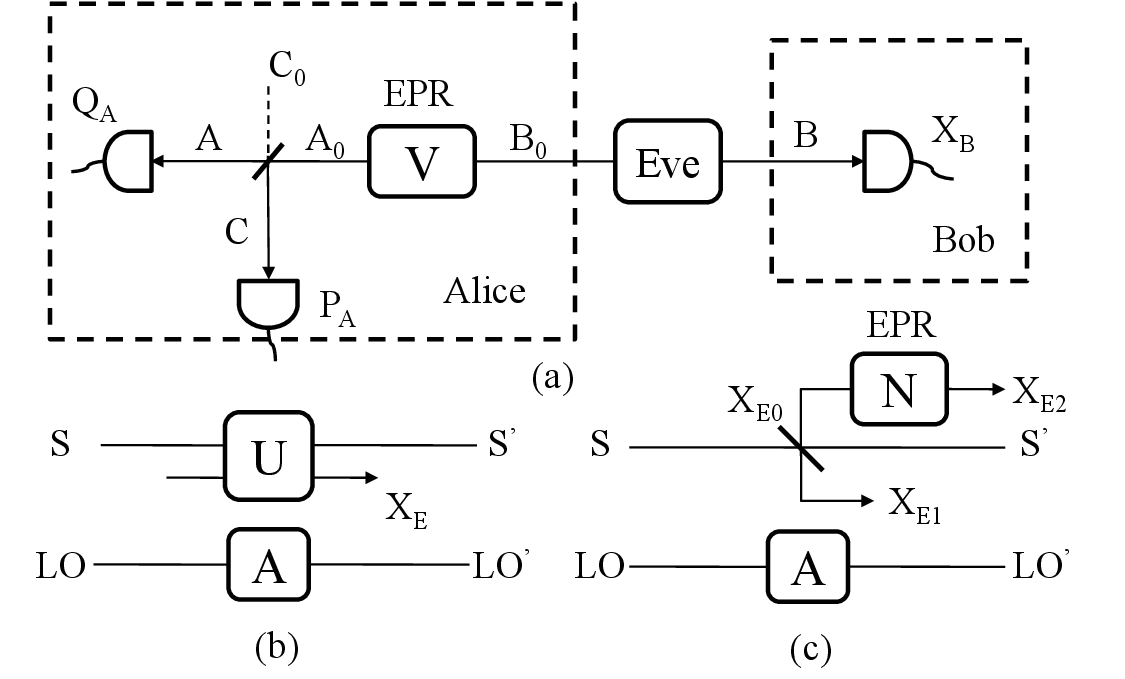}}
 \caption{\label{fig:1}(a) EB scheme, based on Alice heterodyning one half of the modes of the Einstein-Podolsky-Rosen(EPR) \cite{Ein35} state and Bob homodyning the other half, with Eve only attacking the signal. (b) LOIA consisting of attacking the signal beam with a general Gaussian collective attack and attacking LO with intensity attenuation by a non-changing phase attenuator \textbf{A} such as a beam splitter whose transmission is variable. \textbf{U} is a unitary operation and $X_E$ is Eve's intercepted mode. (c) LOIA with \textbf{U} being a beam-splitter transformation and Eve's ancilla modes $X_{E0}$ and $X_{E2}$ being the EPR pairs with variance $N$, and $X_{E1}$ is the intercepted mode. This signal attack is also called an entangling cloner attack.}
\end{figure}

In conventional security analysis, like the EB scheme equivalent to the usual PM implementation depicted in Fig.~\ref{fig:1}(a), LO is not taken into consideration and its intensity is assumed to keep unchanged. However, in practical implementation, Eve could intercept not only the signal beam but also the LO, and she can replace the quantum channel between Alice and Bob with her own perfect quantum channel as shown in Figs.~\ref{fig:1}(b) and \ref{fig:1}(c). In so doing, Eve's attack can be partially hidden by reducing the intensity of LO with a variable attenuator simulating the fluctuation without changing LO's phase, and such an attack can be called a LO intensity attack (LOIA). In the following analysis, we will see that, in the parameter-estimation procedure between Alice and Bob, channel excess noise introduced by Eve can be reduced arbitrarily, even to its being null, just by tuning the LO transmission. Consequently, Alice and Bob would underestimate Eve's intercepted information and Eve could get partial secret keys that Alice and Bob hold without being found under this attack.

Figure \ref{fig:1}(b) describes the LOIA, which consists of attacking the signal beam with a general Gaussian collective attack \cite{Gro05,Gar06,Nav06} and attacking the LO beam with an intensity attenuation by a non-changing phase attenuator \textbf{A}, such as a beam splitter whose transmission is variable. This signal-beam Gaussian collective attack consists of three steps: Eve interacts her ancilla modes with the signal mode by a unitary operation \textbf{U} for each pulse and stores them in her quantum memory, then she makes an optimal collective measurement after Alice and Bob's classical communication. Figure \ref{fig:1}(c) is one practical LOIA with \textbf{U} being a beam-splitter transformation. Its signal attack is also called an entangling cloner attack, which was presented first by Grosshan \cite{Gro03Q} and improved by Weedbrook \cite{Wee10,Wee12A}. In Appendix \ref{sec:Security}, we will demonstrate that with this entangling cloner, Eve can get the same amount of information as that shown in Fig.~\ref{fig:1}(b).

\section{\label{sec:Eskr}Estimation of secret key rate under LOIA}
We analyze a practical CVQKD system with homodyne protocol to demonstrate the effect of LOIA on the secret key rate, and for simplicity we do not give the results of the heterodyne protocol, which is analogous to the homodyne protocol. In the usual PM implementation, Alice prepares a series of coherent states centered on $X_S\in\{Q_S,P_S\}$ with each pulse, and then she sends them to Bob through a quantum channel which might be intercepted by Eve. Here, $Q_S$ and $P_S$, respectively, satisfy a Gaussian distribution independently with the same variance $V_S$ and zero mean. This initial mode prepared by Alice can be described as $\hat{X}_A=X_S+\hat{X}_N$, and $\hat{X}_A\in\{\hat{Q}_A,\hat{P}_A\}$ is the quadrature variable. Here $\hat{X}_N\in\{\hat{Q}_N,\hat{P}_N\}$ describes the quadrature of the vacuum mode. Note that we denote an operator with a hat, while without a hat the same variable corresponds to the classical variable after measurement. So the overall variance of the initial mode prepared by Alice is $V=V_S+1$. When this mode comes to Bob, Bob will get a mode $\hat{X}_B\in\{\hat{Q}_B,\hat{P}_B\}$,
\begin{equation}\label{eq:XB}
\hat{X}_B=\sqrt{T}\hat{X}_A+\sqrt{1-T}\hat{E},
\end{equation}
where $\hat{E}$ describes Eve's mode introduced through the quantum channel whose quadrature variance is $N$. Bob randomly selects a quadrature to measure, and if Eve attenuates the LO intensity during the key distribution, but Bob's outputs still scale with the initial LO intensity, just as in Eq.~(\ref{eq:X'0}), he will get the measurement
\begin{equation}\label{eq:XBm}
X_B^w=\sqrt{\eta}X_B=\sqrt{\eta}(\sqrt{T}X_A+\sqrt{1-T}E),
\end{equation}
where $X_B$ is $Q_B\in\mathbb{R}$ (or equivalently $P_B\in\mathbb{R}$). However, if Bob monitors LO and also scales with the instantaneous intensity value of LO with each pulse, he will get $X_B$ without any loss of course. Note that, for computation simplicity, hereafter we assume the variable transmission rate $\eta$ (or attenuation rate $1-\eta$) of each pulse of LO is the same without loss of generality. Thus the variance of Bob's measurements and conditional variance on Alice's encodings with and without monitoring (in what follows, without monitoring specially indicates that Bob's measurement is obtained just by scaling with the initial LO intensity instead of monitoring instantaneous values, and vice versa) can be given by
\begin{align}
V_B&=TV+(1-T)N\;,\label{eq:VB}\\
V_B^w&=\eta\left[TV+(1-T)N\right]\;,\label{eq:VBM}\\
V_{B|A}&=T+(1-T)N\;, \\
V_{B|A}^w&=\eta\left[T+(1-T)N\right]\;,
\end{align}
where the superscript $w$ indicates ``without monitoring" and all variances are in shot-noise units; the conditional variance is defined as \cite{Gra98}
\begin{equation}\label{eq:CVar}
V_{X|Y}=V(X)-\frac{|\langle XY\rangle|^2}{V(Y)}.
\end{equation}

Hence, the covariance matrix of Alice's and Bob's modes can be obtained as
\begin{align}
\begin{split}
&\gamma_{AB}(V,T,N)=\begin{pmatrix}
                   \gamma_A&    \sigma^T_{AB}\\
                   \sigma_{AB}& \gamma_B
                   \end{pmatrix}\\
                  &\qquad=\begin{pmatrix}
                   V\mathbb{I}&    \sqrt{T(V^2-1)}\sigma_z\\
                   \sqrt{T(V^2-1)}\sigma_z& [TV+(1-T)N]\mathbb{I}
                   \end{pmatrix}\label{eq:rAB},\end{split} \\
&\gamma_{AB}^w=\begin{pmatrix}
                V\mathbb{I}&          \sqrt{\eta T(V^2-1)}\sigma_z\\
                \sqrt{\eta T(V^2-1)}\sigma_z&  \eta[TV+(1-T)N]\mathbb{I}
                \end{pmatrix},
\end{align}
where $\sigma_z=\bigl(\begin{smallmatrix}1 & 0 \\
                                         0 & -1 \end{smallmatrix}\bigr)$ is the Pauli matrix and $\mathbb{I}$ is a unit matrix.

From Eqs.~(\ref{eq:VB}) and (\ref{eq:VBM}) we can derive that the channel transmission and excess noise are $T$, $\varepsilon=(1-T)(N-1)/T$ with monitoring, and $\eta T$, $\varepsilon^w=\varepsilon-\frac{1}{T}(\frac{1}{\eta}-1)$ without monitoring. Hence, by attenuating the LO intensity as Fig.~\ref{fig:1} shows, to make $0<\eta<1$, Eve could arbitrarily reduce $\varepsilon^w$ to zero, thus she will get the largest amount of information permitted by physics. In the following numerical simulation we always make $\varepsilon^w=0$, namely, $\eta(1-T)N=1-\eta T$. Thus, the covariance matrix $\gamma^w_{AB}=\gamma_{AB}(V,\eta T,1)$ and Eve's introducing noise $N$ [in an entangling cloner it is Eve's EPR state's variance as Fig.~\ref{fig:1}(c) shows] should be selected to be
\begin{equation}\label{eq:N}
N=\frac{1-\eta T}{\eta(1-T)}.
\end{equation}
To estimate the secret key rate, without loss of generality, we first analyze the reverse reconciliation then consider the direct reconciliation.

\subsection{Reverse reconciliation}
From Alice and Bob's points of view, the secret key rate for reverse reconciliation with monitoring or not are given, respectively, by
\begin{align}
K_{\text{RR}}=I_{AB}-\chi_{BE}, \label{eq:KRR}\\
K_{\text{RR}}^w=I_{AB}^w-\chi_{BE}^w,\label{eq:KRRm}
\end{align}
where the mutual information between Alice and Bob with and without monitoring are the same, and that is
\begin{equation}\label{eq:IAB}
I_{AB}=\frac{1}{2}\log_2\frac{V_B}{V_{B|A}}=\frac{1}{2}\log_2\frac{V_B^w}{V_{B|A}^w}=I_{AB}^w.
\end{equation}
This is because Bob's measurements in these two cases are just different with a coefficient $\eta$, and they correspond with each other one by one, so they are equivalent according to the data-processing theorem \cite{Kul51,Ziv73}. However, the mutual information between Eve and Bob given by the Holevo bound \cite{Hol99} in these two cases is not identical. As the previous analysis showed, in Bob's point of view, channel transmission and the excess noise estimation are different. But from Eve's point of view, they are identical according to the data-processing theorem because she estimates Bob's measurements in these two cases just by multiplying a coefficient $\eta$. We'll calculate the real information intercepted by Eve first. It can be given by
\begin{equation}\label{eq:XBE0}
\chi_{BE}=S(E)-S(E|B),
\end{equation}
where $S(\cdot)$ is the Von Neumann entropy \cite{Neu55}. For a Gaussian state $\varrho$, this entropy can be calculated by the symplectic eigenvalues of the covariance matrix $\gamma$ characterizing $\varrho$ \cite{Ser04}. To calculate Eve's information, first Eve's system $E$ can purify $AB$ permitted by quantum physics, so that $S(E)=S(AB)$. Second, after Bob's projective measurement, the system $AE$ is pure, so that $S(E|B)=S(A|B)$. Designating $a=V$, $b=TV+(1-T)N$, and $c=\sqrt{T(V^2-1)}$, the symplectic eigenvalues of $\gamma_{AB}$ are given by
\begin{equation}
\lambda_{1,2}=\sqrt{\frac{A\mp\sqrt{A^2-4B^2}}{2}},
\end{equation}
where $A=a^2+b^2-2c^2$ and $B=ab-c^2$. Similarly, the entropy $S(A|B)$ is determined by the symplectic eigenvalue $\lambda_3$ of the covariance matrix $\gamma^{\hat{X}_B}_A$ \cite{Gar07}, namely,
\begin{equation}
\gamma^{\hat{X}_B}_A=\gamma_A-\sigma^T_{AB}(\textbf{X}\gamma_B\textbf{X})^{MP}\sigma_{AB},
\end{equation}
where $\textbf{X}=\bigl(\begin{smallmatrix}1 & 0 \\
                                           0 & 0 \end{smallmatrix}\bigr)$ and $MP$ stands for the Moore-Penrose inverse of a matrix. Then $\lambda_3=\sqrt{\frac{(1-T)NV^2+TV}{TV+(1-T)N}}$, and the Holevo bound reads
\begin{equation}\label{eq:XBE}
\chi_{BE}(V,T,N)=G\left(\frac{\lambda_1-1}{2}\right)+G\left(\frac{\lambda_2-1}{2}\right)-G\left(\frac{\lambda_3-1}{2}\right),
\end{equation}
where $G(x)=(x+1)\log_2(x+1)-x\log_2x$. However, Eve's information estimated by Bob without monitoring is given by
\begin{equation}\label{eq:XBEm}
\chi^w_{BE}=\chi_{BE}(V,\eta T,1).
\end{equation}

By substituting Eqs.~(\ref{eq:XBE}) and (\ref{eq:XBEm}) into Eqs.~(\ref{eq:KRR}) and (\ref{eq:KRRm}), respectively, the secret key rate with and without Bob's monitoring can be obtained. However, the secret key rate in Eq.~(\ref{eq:KRRm}) without monitoring is unsecured in evidence. Eve's interception of partial information from $K^w_{\text{RR}}$ is not detected, in other words, Alice and Bob underestimate Eve's information without realizing it. Actually, the real or unconditionally secure secret key rate $K^w_{\text{RR}}$, which we called a truly secret key rate, should be available by replacing $\chi^w_{BE}$ in Eq.~(\ref{eq:KRRm}) with Eq.~(\ref{eq:XBE}). Note that it is identical with the monitoring secret key rate in Eq.~(\ref{eq:KRR}) due to Eq.~(\ref{eq:IAB}).

\begin{figure}[h]
 \includegraphics[width=\columnwidth]{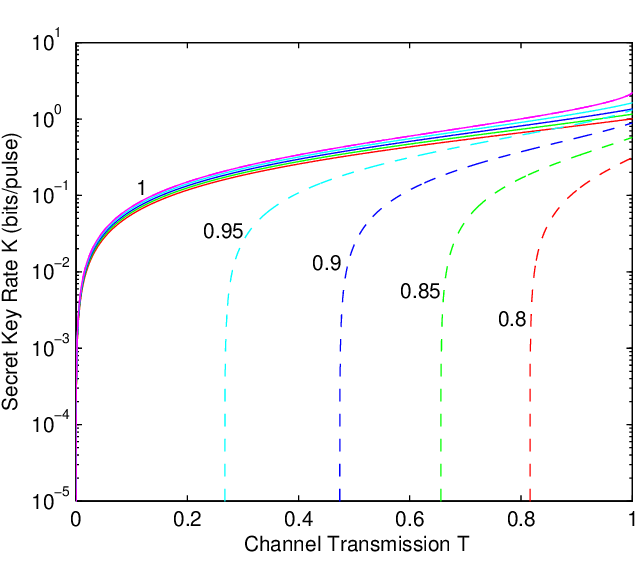}
 \caption{\label{fig:2}(Color online) Reverse reconciliation pseudosecret key rate and the truly secret one vs channel transmission $T$ under LOIA. Solid lines are secret key rate estimated by Bob without monitoring LO intensity and dashed lines are the truly secret ones. Colored lines correspond to the LO transmissions $\eta$ as labeled. Here Alice's modulation variance $V_S=20$.}
\end{figure}

We investigate the secret key rate $K^w_{\text{RR}}$ Bob measured without monitoring and the true one or equivalently monitoring one $K_{\text{RR}}$ for reverse reconciliation under Eve attacking the intensity of LO during key distribution. As Fig.~\ref{fig:2} shows, with various values of transmission of LO that can be controlled by Eve, the truly secret key rate Alice and Bob actually share decreases rapidly over long distances or small channel transmissions.

\begin{figure}[h]
 \scalebox{1}{\includegraphics[width=\columnwidth]{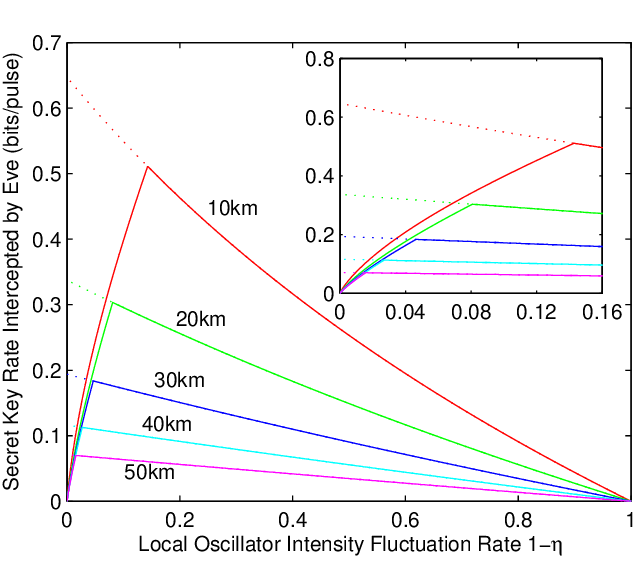}}
 \caption{\label{fig:3}(Color online) Reverse reconciliation pseudosecret key rate (dashed lines) and intercepted one (solid lines) by Eve without being found vs LO intensity fluctuation rate $(1-\eta)$ with different transmission distances. From top to bottom the transmission distances are labeled, and the fiber loss is 0.2 dB/km. Inset is an amplified graph with $(1-\eta)$ between $0\thicksim0.16$. Here Alice's modulation variance $V_S=20$.}
\end{figure}

Additionally, because the mutual information between Alice and Bob with and without monitoring is identical as Eq.(\ref{eq:IAB}) shows, subtracting Eq.~(\ref{eq:KRR}) from Eq.~(\ref{eq:KRRm}) we can estimate Eve's intercepted information $(K^w_{\text{RR}}-K_{\text{RR}})$ which is plotted in Fig.~\ref{fig:3}. We find that Eve could get partial or full secret keys which Alice and Bob hold by controlling the different transmissions of LO. Taking a 20-km transmission distance as an example, surprisingly, just with LO intensity fluctuation or attenuating rate 0.08, Eve is able to obtain the full secret keys for reverse reconciliation without Bob's monitoring the LO.

\subsection{Direct reconciliation}
\begin{figure}[h]
 \scalebox{1}{\includegraphics[width=\columnwidth]{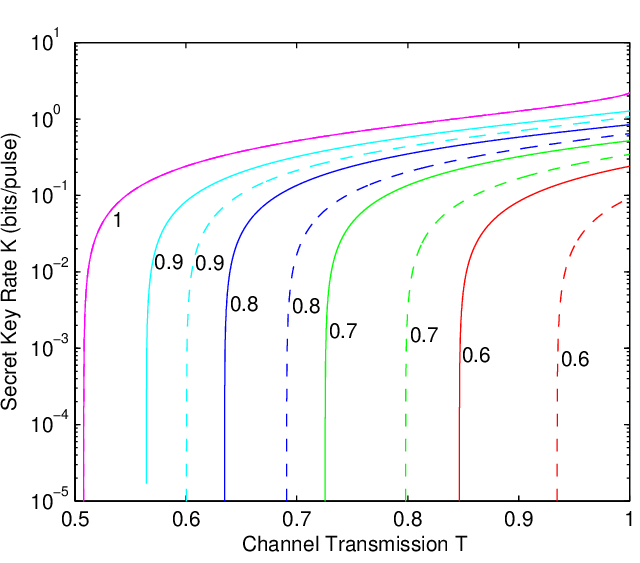}}
 \caption{\label{fig:4}(Color online) Direct reconciliation pseudosecret key rate and truly secret one vs channel transmission $T$ under LOIA. Solid lines are secret key rates estimated by Bob without monitoring LO intensity and dashed lines are the truly secret ones. Colors correspond to the LO transmissions as labeled. Here Alice's modulation variance $V_S=20$.}
\end{figure}

We now calculate the secret key rate for direct reconciliation, which is a little more complicated, and we investigate the effect of LO intensity attack by Eve on CVQKD. The secret key rate estimated by Bob with and without LO monitoring is given, respectively, by
\begin{align}
K_{\text{DR}}=I_{AB}-\chi_{AE},\label{eq:KDR}\\
K_{\text{DR}}^w=I_{AB}^w-\chi_{AE}^w.\label{eq:KDRm}
\end{align}
Note that we have already calculated $I_{AB}$ and $I^w_{AB}$ in Eq.(\ref{eq:IAB}) and they are identical for direct and reverse reconciliation. For Eve we have
\begin{equation}\label{eq:XAE0}
\chi_{AE}=S(E)-S(E|A),
\end{equation}
where $S(E)=S(AB)$ has been already computed in the previous section, and $S(E|A)=S(BC|A)$ using the fact that after Alice's projective measurement on modes $A_0$ and $C_0$ obtaining $Q_A$ in the EB scheme shown in Fig.~\ref{fig:1}(a), the system $BCE$ is pure. To calculate $S(BC|A)$, we have to compute the symplectic eigenvalues of covariance matrix $\gamma^{\hat{X}_A}_{BC}$, which is obtained by
\begin{equation}
\gamma^{\hat{X}_A}_{BC}=\gamma_{BC}-\sigma^T_{BCA}(\textbf{X}\gamma_A\textbf{X})^{MP}\sigma_{BCA},
\end{equation}
where $\gamma_{BC}$ and $\sigma_{BCA}$ can be read in the decomposition of the matrix
\begin{equation}
\gamma_{BCA}=\begin{pmatrix}
             \gamma_{BC}&     \sigma^T_{BCA}\\
             \sigma_{BCA}&    \gamma_A
             \end{pmatrix},
\end{equation}
which is available by elementary transformation of the matrix [see Fig.~\ref{fig:1}(a)] \cite{Lod07,Gar07}
\begin{equation}
\gamma_{ACB}=\big(S^{BS}_{A_0C_0}\oplus\mathbb{I}_B\big)^T\gamma_{A_0C_0B}\big(S^{BS}_{A_0C_0}\oplus\mathbb{I}_B\big),
\end{equation}
where $\mathbb{I}_B$ is a unit matrix. It is obtained by applying a homodyne detection on mode A after mixing $A_0$ and $C_0$ with a balanced beam-splitter transformation $(S^{\text{BS}}_{A_0C_0})$. The matrix $\gamma_{A_0C_0B}=\gamma_{A_0B}\oplus\gamma_{C_0}$ and $\gamma_{A_0B}$ actually is $\gamma_{AB}$ in Eq.~(\ref{eq:rAB}), $\gamma_{C_0}$ is a unit matrix. So we can get
\begin{equation}
\gamma^{\hat{X}_A}_{BC}\negthickspace=\negthickspace\begin{pmatrix}
                        b-c^2/(a+1)& 0& \sqrt{2}c/(a+1)& 0\\
                        0& b& 0& -c/\sqrt{2}\\
                        \sqrt{2}c/(a+1)& 0& 2a/(a+1)& 0\\
                        0& -c/\sqrt{2}& 0& (a+1)/2
                        \end{pmatrix}
\end{equation}
and the symplectic eigenvalues of it
\begin{equation}
\lambda_{4,5}=\sqrt{\frac{C\mp\sqrt{C^2-4D}}{2}},
\end{equation}
where $C=\frac{a+bB+A}{a+1}$ and $D=\frac{B(b+B)}{a+1}$. The Holevo bound then reads
\begin{align}
&\chi_{AE}(V,T,N)=\sum^2_{i=1}G\left(\frac{\lambda_i-1}{2}\right)-\sum^5_{j=4}G\left(\frac{\lambda_j-1}{2}\right),\label{eq:XAE}\\
&\chi^w_{AE}=\chi_{AE}(V,\eta T,1).\label{eq:XAEm}
\end{align}
Substituting Eqs.~(\ref{eq:XAE}) and (\ref{eq:XAEm}) into Eqs.~(\ref{eq:KDR}) and (\ref{eq:KDRm}), respectively, the secret key rates in these two cases are obtained. In Fig.~\ref{fig:4}, we plotted them for channel transmission $T$ with various values of $\eta$ and find that the difference between the pseudosecret key rate with Bob not monitoring LO and the truly secret one is still increasing with the channel transmission $T$ becoming smaller.

\begin{figure}[h]
 \scalebox{1}{\includegraphics[width=\columnwidth]{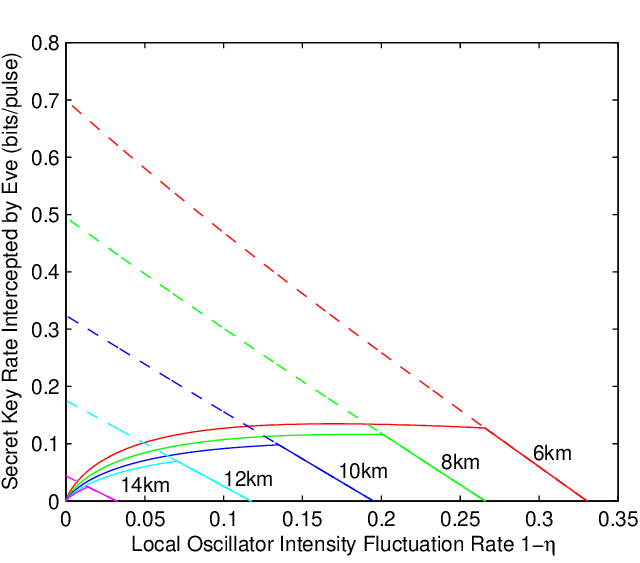}}
 \caption{\label{fig:5}(Color online) Direct reconciliation pseudosecret key rate (dashed lines) and intercepted one (solid lines) by Eve without being found vs LO intensity fluctuation rate $(1-\eta)$ with different transmission distances. From top to bottom the transmission distances are labeled, and the fiber loss is 0.2 dB/km. Here Alice's modulation variance $V_S=20$.}
\end{figure}

For Eve, when Bob does not monitor LO, she will get the partial or total secret key rate $(K^w_{\text{DR}}-K_{\text{DR}})$ without being found by subtracting Eq.~(\ref{eq:KDR}) from Eq.~(\ref{eq:KDRm}) when she reduces the intensity of LO. In Fig.~\ref{fig:5}, we plotted the pseudosecret key rate for direct reconciliation and the mutual information overestimated by Alice and Bob. We find that for short distance communication (less 15 km or 3 dB limit), a small fluctuation of LO intensity could still hide Eve's attack partially or totally.

Note that in the above estimation we assume each pulse's transmission rate $\eta$ (or attenuation rate $1-\eta$) is identical. However, when $\eta$ is different for each pulse (Eve simulates the fluctuation of LO to hide her dramatic attack on LO), Eve still could intercept as much as or even more secret key rates than above for reverse and direct reconciliation, as long as the largest value of $\eta$ among all pulses (or approximately most pulse transmission rates) is smaller than the above constant value.

\section{\label{sec:Discussion}Discussion and conclusion}
Our analysis shows that reverse reconciliation is more sensitive than direct reconciliation about the fluctuation of LO intensity, and, even with a small attenuation of LO intensity, Eve can get full secret keys but not be found. This is consistent with the fact that channel excess noise has a more severe impact on reverse reconciliation than on direct reconciliation. Of course, when the intensity of LO fluctuates above the initial calibrated value (i.e., $\eta>1$), Eve could not get any secret keys, but Alice and Bob would overestimate Eve's intercepted information due to the overestimation of channel excess noise. However, when LO fluctuates around the initial calibrated value, how to quantify Eve's information is still an open question, because the distribution of the fluctuation of LO (or $\eta$) is not a normal distribution and unclear for Alice and Bob due to Eve's arbitrary manipulation. But in this circumstance, Eve still could intercept partial secret keys if she increases the channel excess noise of one part of the signal pulses when she controls $\eta<1$ and decreases it for the other part when controlling $\eta>1$, i.e., making the overall estimated excess noise by Alice and Bob lower than the real one. Remarkably, LO intensity fluctuation opens a loophole for Eve to attack the practical system, especially in the case of communication with low channel transmission or over long distance.

Consequently, in the practical implementation of CVQKD, we must monitor the LO fluctuation carefully and in particular scale the measurements with instantaneous intensity values of LO. Alternatively, we can also scale with the lowest intensity value of LO if the fluctuations are very small, but it will estimate the secret key rate pessimistically thus leading to the reduction of the efficiency of the key distribution. However, we cannot use the average intensity value of LO to normalize the measurements as most current implementations do, because it still could overestimate the secret key rate for Alice and Bob. Additionally, for reverse reconciliation communication over long distance, very small fluctuation of LO might compromise the secret key rate completely, which presents a big challenge for accurately monitoring LO intensity.

Finally, we point out that in this paper we do not consider the imperfections of BHD such as detection efficiency, electronic noise, and incomplete subtraction, which may make LO intensity fluctuation have a more severe impact on estimating the secret key rate for Alice and Bob.

In conclusion, we have analyzed the effect of LO intensity fluctuation on the secret key rate estimation of Alice and Bob for reverse and direct reconciliation. Incredibly, Bob's estimation of the secret key rate will be compromised severely without monitoring LO or if his measurements do not scale with LO instantaneous intensity values even with monitoring but just discard large fluctuation pulses like in \cite{YM11}. Furthermore, we have shown that Eve could hide her attack partially by reducing the intensity of LO and even could steal the total secret keys Alice and Bob share without being found by a small attenuation of LO intensity, especially for reverse reconciliation. Finally, we have also briefly discussed the monitoring of LO and pointed out that it would be a challenge for highly accurate monitoring.


\begin{acknowledgments}
This work is supported by the National Natural Science Foundation of China, Grant No. 61072071. L.-M.L. is supported by the Program for New Century Excellent Talents. X.-C.M. is supported by the Hunan Provincial Innovation Foundation for Postgraduates. X.-C.M. and M.-S.J. acknowledge support from NUDT under Grant No. kxk130201.
\end{acknowledgments}

\appendix
\section{Security estimation by Entangling Cloner under LOIA}\label{sec:Security}
In this Appendix, we calculate the Holevo bound obtained by Eve for direct and reverse reconciliation using Weedbrook's entangling cloner model \cite{Wee10,Wee12A}, and then give the secret key rate shared by Alice and Bob under LOIA. We begin the analysis by calculating the Von Neumann entropy of Eve's intercepting state first.

As Fig.~\ref{fig:1}(c) shows, the entangling cloner consists of Eve replacing the Gaussian quantum channel between Alice and Bob with a beam splitter of transmission $T$ and an EPR pair of variance $N$. Half of the EPR pair mode $E_0$ is mixed with Alice's mode in the beam splitter and is sent to Bob to match the noise of the real channel by tuning N. The other half mode $E_2$ is kept by Eve to reduce the uncertainty on one output of the beam splitter, the mode $E_1$, which can be read as
\begin{equation}
\hat{X}_{E_1}=-\sqrt{1-T}\hat{X}_A+\sqrt{T}\hat{X}_{E_0},
\end{equation}
where $\hat{X}_{E_0}$ is the quadrature of mode $E_0$. Thus, the variance of mode $E_1$ is given by
\begin{equation}
V_{E_1}=(1-T)V+TN,
\end{equation}
and the conditional variance $V_{E_1|A}$ can be calculated as, using Eq.~(\ref{eq:CVar}),
\begin{equation}
V_{E_1|A}=(1-T)+TN.
\end{equation}
Hence, Eve's covariance matrix can be obtained as
\begin{equation}
\gamma_E(V,V)\negthickspace=\negthickspace\begin{pmatrix}
              \gamma_{E_1}& \sigma^T_{E_1E_2}\\
              \sigma_{E_1E_2}& \gamma_{E_2}
              \end{pmatrix}\negthickspace=\negthickspace
              \begin{pmatrix}
              \text{diag}(V_{E_1},V_{E_1})& Z_{E_1\!E_2}\sigma_z\\
              Z_{E_1\!E_2}\sigma_z& N\mathbb{I}
              \end{pmatrix},
\end{equation}
where $Z_{E_1E_2}=\sqrt{T(N^2-1)}$ and the notation $\text{diag}(,)$ stands for a matrix with the arguments on the diagonal elements and zeros everywhere else. The symplectic eigenvalues of this covariance matrix are given by
\begin{equation}
\lambda_{1,2}=\sqrt{\frac{\Delta\mp\sqrt{\Delta^2-4D}}{2}},
\end{equation}
where $\Delta=V^2_{E_1}+N^2-2Z^2_{E_1E_2}$, and $D=(V_{E_1}N-Z^2_{E_1E_2})^2$. Hence, the Von Neumann entropy of Eve's state is given by
\begin{equation}\label{eq:SE}
S(E)=G\left(\frac{\lambda_1-1}{2}\right)+G\left(\frac{\lambda_2-1}{2}\right),
\end{equation}

\subsection{Direct reconciliation}
For the direct reconciliation protocol of CVQKD, the Holevo bound between Eve and Alice is given by Eq.~(\ref{eq:XAE0}), where $S(E)$ has been calculated by Eq.~(\ref{eq:SE}). $S(E|A)$ can be obtained by the conditional covariance matrix
\begin{equation}
\gamma^{X_A}_E=\gamma_E(V=1,V),
\end{equation}
and its symplectic eigenvalues are given by
\begin{equation}
\lambda_{3,4}=\sqrt{\frac{A\mp\sqrt{A^2-4B}}{2}},
\end{equation}
where $A=V_{E_1|A}V_{E_1}+N^2-2Z^2_{E_1E_2}$, and $B=(V_{E_1|A}N-Z^2_{E_1E_2})(V_{E_1}N-Z^2_{E_1E_2})$. Thus, the conditional entropy is
\begin{equation}\label{eq:SEA}
S(E|A)=G\left(\frac{\lambda_3-1}{2}\right)+G\left(\frac{\lambda_4-1}{2}\right).
\end{equation}
Substituting Eqs.~(\ref{eq:SE}) and (\ref{eq:SEA}) into Eq.~(\ref{eq:XAE0}), we can get the mutual information between Alice and Eve,
\begin{equation}\label{eq:XAE1}
\chi_{AE}(V,T,N)=S(E)-S(E|A).
\end{equation}
Under LOIA, Bob's estimation of the Holevo bound without monitoring LO intensity then reads, using Eq.~(\ref{eq:XAE1}),
\begin{equation}\label{eq:XAEm1}
\chi^w_{AE}=\chi_{AE}(V,\eta T,1).
\end{equation}
With Eqs.~(\ref{eq:XAE1}) and (\ref{eq:XAEm1}), the secret key rates in Eqs.~(\ref{eq:KDR}) and (\ref{eq:KDRm}) then can be calculated respectively, and the calculation numerically demonstrates that they are perfectly consistent with the Fig.~\ref{fig:4}.

\subsection{Reverse reconciliation}
The calculation of the Holevo bound between Eve and Bob for reverse reconciliation is a bit more complicated. Using Eq.~(\ref{eq:XBE0}), we only need to calculate the conditional entropy $S(E|B)$, which is determined by the symplectic eigenvalues $\lambda_{4,5}$ of the covariance matrix $\gamma^{\hat{X}_B}_E$,
\begin{equation}
\gamma^{\hat{X}_B}_E=\gamma_E-\sigma^T_{E_1E_2B}(\textbf{X}\gamma_B\textbf{X})^{MP}\sigma_{E_1E_2B},
\end{equation}
where $\sigma_{E_1E_2B}=(\langle\hat{X}_{E_1}\hat{X}_B\rangle \mathbb{I}, \langle\hat{X}_{E_2}\hat{X}_B\rangle \sigma_z)=(Z_{E_1B} \mathbb{I}, Z_{E_2B} \sigma_z)$ and $Z_{E_1B}=\sqrt{T(1-T)}(N-V)$, $Z_{E_2B}=\sqrt{1-T}\sqrt{N^2-1}$. Then, $\gamma^{\hat{X}_B}_E$ can be recast as
\begin{equation}
\gamma^{\hat{X}_B}_E=\begin{pmatrix}
                     F& H^T\\
                     H& G
                     \end{pmatrix}\tag{A12$'$}
\end{equation}
where \\
$F=\text{diag}\left(V_{E_1}\negthickspace-\negthickspace\frac{Z^2_{E_1B}}{V_B}, V_{E_1}\right)$, $G=\text{diag}\left(N\negthickspace-\negthickspace\frac{Z^2_{E_2B}}{V_B}, N\right)$, \\
and\\
    $H=\text{diag}\left(Z_{E_1E_2}-\frac{Z_{E_1B}Z_{E_2B}}{V_B}, -Z_{E_1E_2}\right)$. \\
Hence, its symplectic eigenvalues are given by
\begin{equation}
\lambda_{5,6}=\sqrt{\frac{C\mp\sqrt{C^2-4D'}}{2}},
\end{equation}
where $C=det(F)+det(G)+2det(H)$, $D'=\text{det}(\gamma^{\hat{X}_B}_E)$, and $\text{det}(\cdot)$ is the determinant of a matrix. So, we get the conditional entropy
\begin{equation}
S(E|B)=G\left(\frac{\lambda_5-1}{2}\right)+G\left(\frac{\lambda_6-1}{2}\right),
\end{equation}
and then the Holevo bound
\begin{equation}\label{eq:XBE1}
\chi_{BE}(V,T,N)=S(E)-S(E|B).
\end{equation}
Consequently, without monitoring LO intensity, Alice and Bob will give Eve the Holevo bound
\begin{equation}\label{eq:XBEm1}
\chi^w_{BE}=\chi_{BE}(V,\eta T,1).
\end{equation}
Substituting Eqs.~(\ref{eq:XBE1}) and (\ref{eq:XBEm1}) into Eqs.~(\ref{eq:KRR}) and (\ref{eq:KRRm}), respectively, the secret key rates with and without Bob's monitoring can be obtained, and for channel transmission $T$ with various values of $\eta$, they are numerically demonstrated to be perfectly consistent with Fig.~\ref{fig:2}, too. Hence, it also indirectly confirms that either for direct or reverse reconciliation, the entangling cloner could reach the Holevo bound against the optimal Gaussian collective attack.

%


\begin{thebibliography}{99}

\bibitem{Wee12R} C. Weedbrook, S. Pirandola, R. Garc{\'{\i}}a-Patr{\'{o}}n, N. J. Cerf, T. C. Ralph, J. H. Shapiro, and S. Lloyd, Rev. Mod. Phys. \textbf{84}, 621 (2012).

\bibitem{Gro03N} F. Grosshans, G. V. Asschee, J. Wenger, R. Brouri, N. Cerf, and P. Grangier, Nature (London) \textbf{421}, 238 (2003).

\bibitem{Lev09N} A. Leverrier, E. Karpov, P. Grangier, and N. J. Cerf, New J. Phys. \textbf{11}, 115009 (2009).

\bibitem{Lev09L} A. Leverrier and P. Grangier, Phys. Rev. Lett. \textbf{102}, 180504 (2009).

\bibitem{Xua09} Q. D. Xuan, Z. Zhang, and P. L. Voss, Opt. Exp. \textbf{17}, 24244 (2009).

\bibitem{Lod07} J. Lodewyck \textit{et al.}, Phys. Rev. A \textbf{76}, 042305 (2007).

\bibitem{Qi07} B. Qi, L. L. Huang, L. Qian, and H. K. Lo, Phys. Rev. A \textbf{76}, 052323 (2007).

\bibitem{Fos09} S. Fossier, E. Diamanti, T. Debuisschert, A. Villing, R. Tualle-Brouri, and P. Grangier, New J. Phys. \textbf{11}, 045023 (2009).

\bibitem{Jou12} P. Jouguet \textit{et al.}, Opt. Exp. \textbf{20}, 14031 (2012).

\bibitem{Zha09} Z. Zhang and P. L. Voss, Opt. Exp. \textbf{17}, 12090 (2009).

\bibitem{She10} Y. Shen, H.-X. Zou, L. Tian, P.-X Chen, and J.-M Yuan, Phys. Rev. A \textbf{82}, 022317 (2010).

\bibitem{Gro05} F. Grosshans, Phys. Rev. Lett. \textbf{94}, 020504 (2005).

\bibitem{Gar06} R. Garc{\'{\i}}a-Patr{\'{o}}n and N. J. Cerf, Phys. Rev. Lett. \textbf{97}, 190503 (2006).

\bibitem{Nav06} M. Navascu\'{e}s, F. Grosshans, and A. Ac\'{\i}n, Phys. Rev. Lett. \textbf{97}, 190502 (2006).

\bibitem{Has08} H. Haseler, T. Moroder, and N. L\"{u}tkenhaus, Phys. Rev. A \textbf{77}, 032303 (2008).

\bibitem{YM11} Y.-M. Chi \textit{et al.}, New J. Phys. \textbf{13}, 013003 (2011).

\bibitem{Ray95} M. G. Raymer, J. Cooper, H. J. Carmichael, M. Beck, and D. T. Smithey, J. Opt. Soc. Am. B \textbf{12}, 1801 (1995).

\bibitem{foot} In this paper, LO intensity fluctuation indicates the deviation of each pulse's intensity from the initial calibrated value during the key distribution. It does not mean the quantum fluctuation of each pulse itself, because LO is a strong classical beam whose quantum fluctuation is very small relative to itself and can be neglected.

\bibitem{Ein35} A. Einstein, B. Podolsky, and N. Rosen, Phys. Rev. \textbf{47}, 777(1935).

\bibitem{Gro03Q} F. Grosshans, N. J. Cerf, J. Wenger, R. Tualle-Brouri, and P. Grangier, Quantum Inf. Comput.\textbf{ 3}, 535 (2003).

\bibitem{Wee10} C. Weedbrook, S. Pirandola, S. Lloyd, and T. C. Ralph, Phys.Rev. Lett. \textbf{105}, 110501 (2010).

\bibitem{Wee12A} C. Weedbrook, S. Pirandola, and T. C. Ralph, Phys. Rev. A \textbf{86}, 022318 (2012).

\bibitem{Gra98} P. Grangier, J. A. Levenson, and J. P. Poizat, Nature (London) \textbf{396}, 537 (1998).

\bibitem{Kul51} S. Kullback and R. Leibler, Ann. Math. Statist. \textbf{22}, 79 (1951).

\bibitem{Ziv73} J. Ziv and M. Zakai, IEEE Trans. Inf. Theory  \textbf{19}, 275 (1973).

\bibitem{Hol99} A. S. Holevo, M. Sohma, and O. Hirota, Phys. Rev. A \textbf{59}, 1820 (1999).

\bibitem{Neu55} J. von Neumann, {\it Mathematical Foundation of Quantum Mechanics} (Princeto University Press, Princeton, NJ, 1955).

\bibitem{Ser04} A. Serafini, F. Illuminati, and S. De Siena, J. Phys. B \textbf{37}, L21 (2004); G. Adesso, A. Serafini, and F. Illuminati, Phys. Rev. A \textbf{70}, 022318 (2004).

\bibitem{Gar07} R. Garc{\'{\i}}a-Patr{\'{o}}n, Ph. D. thesis, Universit${\rm \acute{e}}$ Libre de Bruxelles, 2007.


\end{thebibliography}
\end{document}